\documentclass[onecolumn, 12pt]{IEEEtran}
\usepackage{amssymb,amsmath}
\usepackage[dvips]{graphicx}
\usepackage{amsfonts}
\usepackage{bm}
\usepackage{subfigure}
\usepackage{booktabs,multirow}
\usepackage{color}
\usepackage{stfloats}
\usepackage{setspace}
\usepackage{algorithm} %format of the algorithm
\usepackage{algorithmic} %format of the algorithm
\usepackage{multirow}                             %multirow                             for                             format
\usepackage{amsmath}
\usepackage{xcolor}
\usepackage{cite}
\usepackage{amsthm}

\IEEEoverridecommandlockouts

\newtheorem{theorem}{Theorem}
\newtheorem{remark}{Remark}

\newtheorem{definition}{Definition}

\begin{document}

\title{Capacity-Aware Edge Caching in Fog Computing Networks}

\author{Qiang~Li,~\IEEEmembership{Member,~IEEE}, Yuanmei~Zhang, Yingyu~Li, \\
Yong~Xiao,~\IEEEmembership{Senior Member,~IEEE}, and Xiaohu Ge,~\IEEEmembership{Senior Member,~IEEE}

\thanks{Qiang~Li, Yuanmei~Zhang, Yingyu Li (corresponding author), Yong Xiao, and Xiaohu Ge are with Huazhong University of Science and Technology, 430074 P. R. China. Emails: \{qli\_patrick, ym\_zhang, liyingyu, yongxiao\}@hust.edu.cn, xhge@mail.hust.edu.cn.}

%\thanks{The authors would like to acknowledge the support from National Key R\&D Program of China (2016YFE0133000): EU-China study on IoT and 5G (EXCITING-723227).}
}
\maketitle

\begin{abstract}
This paper studies edge caching in fog computing networks, where a capacity-aware edge caching framework is proposed by considering both the limited fog cache capacity and the connectivity capacity of base stations (BSs). By allowing cooperation between fog nodes and cloud data center, the average-download-time (ADT) minimization problem is formulated as a multi-class processor queuing process. We prove the convexity of the formulated problem and propose an Alternating Direction Method of Multipliers (ADMM)-based algorithm that can achieve the minimum ADT and converge much faster than existing algorithms. Simulation results demonstrate that the allocation of fog cache capacity and connectivity capacity of BSs needs to be balanced according to the network status. While the maximization of the edge-cache-hit-ratio (ECHR) by utilizing all available fog cache capacity is helpful when the BS connectivity capacity is sufficient, it is preferable to keep a lower ECHR and allocate more traffic to the cloud when the BS connectivity capacity is deficient.
\end{abstract}
%Furthermore, a near-optimal heuristic caching algorithm is proposed which indicates that

%This paper studies edge caching in fog computing networks, where a set of distributed fog nodes have been deployed to cache contents and directly serve the requesting users upon a cache hit. In contrast to most existing works focusing on optimizing content placement in the cache of each fog node while ignoring the provision of the cached content, we propose a traffic-aware edge caching framework taking into consideration of both the cache capacity limit and the content provision capability of each fog node. Then the minimization of the average download time per content request is formulated as a multi-class processor sharing queue problem by using M/M/1 queue model. A distributed algorithm based on Alternating Direction Method of Multipliers (ADMM) is then proposed for searching the optimal cache content placement at each fog node. A near-optimal heuristic caching algorithm is proposed to provide insights into the relationship between the cache capacity limit and the content provision rate. Simulation results demonstrate the maximization of the edge-cache-hit-ratio (ECHR) by utilizing all available cache capacity may incur additional queuing delay at the fog node. Also, when the content request arrival rate is relatively high, it is preferable to keep a lower ECHR and allocate more traffic to the cloud.

%cache capacity & provision rate相互制约

%model the content requests arrived at each fog node as a Poisson process and

\begin{IEEEkeywords}
Edge caching, fog computing, average-download-time, M/M/1 queue, ADMM.
\end{IEEEkeywords}

\IEEEpeerreviewmaketitle

\vspace{-0.3cm}

%%%%%%%%%%%%%%%%%%%%%%%%%%%%%%%%%%%%%%%%%%%%%%%%%%
\section{Introduction}
%%%%%%%%%%%%%%%%%%%%%%%%%%%%%%%%%%%%%%%%%%%%%%%%%%%

\vspace{-0.1cm}
Fog computing is a virtualized network architecture that relies on a large number of low-cost and often decentralized servers, called fog nodes, to perform caching and computing at the edge of networks. It has been considered as a promising solution for supporting delay-sensitive and bandwidth-consuming applications such as AR/VR and video streaming in 5G and beyond networking systems \cite{fog_groups}. By caching popular contents that are frequently requested by users at their proximal fog nodes, edge caching has attracted significant interest in both industry and academia due to its capability to reduce content delivery latency for users and alleviate traffic congestion in the backhaul and backbone networks \cite{fog_groups, cache_air,Ge2013Energy}.

Due to the cache capacity limits of fog nodes, most existing works focus on maximizing the amounts of popular contents that can be stored in each node. To characterize the efficiency of caching strategies, edge-cache-hit ratio (ECHR) has been proposed as an important performance metric to evaluate the probability of users' content requests being delivered at the network edge. It is commonly believed in \cite{fogcaching_maxhit, D2D_ECHR, fog_cluster_maxhit, D2D_TZ} that the maximization of the ECHR is helpful to improve the users' quality-of-experience (QoE) as well as the utilization efficiency of cache-enabled networks.

However, we observe that maximizing the ECHR can only improve users' QoE if the contents cached by fog nodes can be successfully delivered to the requesting users in a timely manner. Unfortunately, the last-mile connectivity between users and base stations (BSs) is usually through wireless links with limited capacities \cite{Ge2017Heterogeneous,Ge2011Capacity}. When the users' content requesting rate exceeds a certain threshold, it may result in intolerable queueing delay in edge content delivery. In this case, simply maximizing ECHR may not be an ideal choice. Although there have been numerous works focusing on optimizing either the cache content placement mechanism of fog nodes or strategies to alleviate the congestion in the radio access networks between users and BSs, there is still a lack of a comprehensive solution that optimizes edge caching by jointly considering the cache capacity, the popularity of content, the users' content requesting rate, and the capacity of the last mile connection between BSs and users.

This motivates the work in this paper where we consider content offloading between cloud data center (CDC) and fog nodes with limited last-mile channel capacities. We propose a capacity-aware edge caching (CAEC) framework to cache contents at fog nodes according to both the limited fog cache capacity and the connectivity capacity between BSs and users. The main contributions are summarized as follows:
\vspace{-0.1cm}
\begin{itemize}
\item We have shown that the edge content delivery can be considered as a special case of a multi-class processor queueing process. We then formulate the proposed CAEC scheme as an optimization problem of the average-download-time (ADT), which has non-negative semi-definite Hessian matrix and can be proved to be convex.

\item In order to determine the best CAEC strategy under different fog cache capacity limits and last mile connectivity capacities of BSs, we propose an Alternating Direction Method of Multipliers (ADMM)-based algorithm that can converge to the global optimal solution much faster compared to traditional methods. In addition, a simple heuristic caching method is proposed that is shown to achieve a near-optimal performance.

%with a linear convergence rate,  can converge

\item Extensive simulation has been performed to demonstrate the performance of the proposed CAEC framework and comparison with traditional ECHR maximization solution. Our result shows that it is not always beneficial to maximize the ECHR, especially when the users' requesting rate is high yet the BS connectivity capacity is low.
%Specifically, CAEC outperforms the ECHR maximization solution by over 6\%.

\end{itemize}

\vspace{-0.3cm}
%%%%%%%%%%%%%%%%%%%%%%%%%%%%%%%%%%%%%%%%%%%%%%%%%%
\section{System Model}
%%%%%%%%%%%%%%%%%%%%%%%%%%%%%%%%%%%%%%%%%%%%%%%%%%

%evolved packet core vs orchestrator
% also referred to as Fog-Access Point (F-AP) \cite{fog_cluster_maxhit},

\begin{figure}[t]
\begin{center}
\vspace{-0.2cm}
{\includegraphics[width=2.6in]{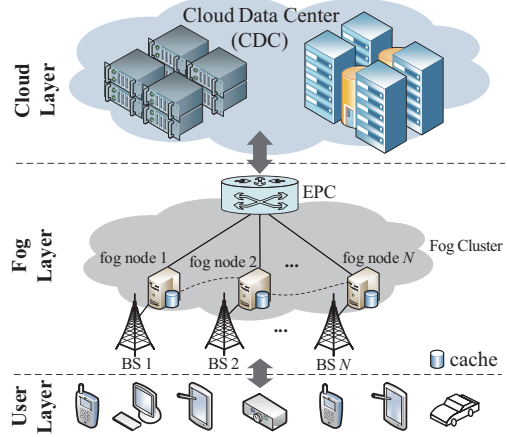}}
\end{center}\vspace{-0.3cm}
\caption{An illustration of the cache-enabled fog computing network.}\label{system}
\vspace{-0.8cm}
\end{figure}

\vspace{-0.2cm}
In this paper, we consider a fog computing-enabled wireless network with a number of $N$ BSs that are carefully deployed throughout the service area to offer wireless services with limited connectivity capacities to their users, as shown in Fig. \ref{system}. Without loss of generality, we assume that each user can only request content files from a finite-sized library ${\mathcal{F}}=\{1,2,...,F\}$ \cite{fogcaching_latency2, fog_caching4, segement}. Each BS is attached with a fog node with limited cache capacity \cite{cache_air}, which is also referred to as Fog-Access Point (F-AP) in \cite{fog_groups}, \cite{fog_cluster_maxhit}, \cite{fogcaching_latency2}. Thus only a finite number of popular contents can be cached for local access of its users. Other contents must be fetched from the remote CDC that can be exclusively owned by a content provider or rented from cloud service providers.

Meanwhile, each user must be associated to a BS before obtaining the requested contents from fog nodes or CDC. Since fog nodes are installed inside the access network infrastructure, they can be connected with each other via high-capacity fiber links \cite{fog_groups}. Thus the $N$ fog nodes can form a fog cluster $\mathcal{N}=\{1,...,N\}$ with the coordination of the evolved packet core (EPC), which is also capable of coordinating the cache content placement among fog nodes \cite{cache_air}, \cite{fog_caching4}. In this case, within a cluster of fog nodes different sets of contents can be stored and shared with each other when necessary.

%As mentioned earlier, major MNOs are deploying fog computing infrastructure for supporting latency sensitive service to their users. To make sure  offer prioritized service to a group of users as shown in Fig. \ref{system}.
%Each user can request
%consisting of four elements: cloud service provider, fog network providers, wireless service providers, and user equipments.
%We assume fog nodes are deployed by MNOs for offering local caching services to the
%As shown in Fig. \ref{system}, a generic cache-enabled fog computing network consists of three layers: user layer, fog layer and cloud layer. User layer consists of a large number of end users, each is associated with a fog node based on its physical proximity, channel conditions and prior service agreements between users and fog nodes \cite{Xiao_JSAC}.

%$N$ closely located fog nodes in fog layer can form a cluster $\mathcal{N}=\{1,...,N\}$ with the coordination of orchestrator, which is also capable of coordinating the cache content placement among fog nodes . In the cloud layer, the content will be provisioned to end users with relatively large latency compared to the fog layer.

\vspace{-0.3cm}
\subsection{Cache Content Placement}
\vspace{-0.2cm}
With the popularity of high-definition videos, content size can be large and it is preferable to cache a portion rather than the entire content in a fog node \cite{fog_groups}, \cite{segement}. We assume that each content can be split into multiple segments and define $P(i,f)$ as the portion of content $f$ cached in fog node $i$.

%, for $0\leq P(i,f) \leq 1$, $i\in\mathcal{N}$ and $f\in {\mathcal{F}}$
%\begin{eqnarray}\label{matrix}
%\mathbf{P} = {\left[ {\begin{array}{*{20}{c}}
%{P(1,1)}&{P(1,2)}& \cdots &{P(1,F)}\\
% \vdots & \vdots & \ddots & \vdots \\
%{P(N,1)}&{P(N,2)}& \cdots &{P(N,F)}
%\end{array}} \right]_{N \times F}},
%\end{eqnarray}
\vspace{-0.2cm}
\begin{definition}
For a cluster of $N$ fog nodes $\mathcal{N}=\{1,...,N\}$, we define a cache content placement matrix $\mathbf{P} \in \mathbb{R}^{N\times F}$, whose entry $P(i,f)$ satisfies the following constraints:
\begin{subequations}
\begin{eqnarray}
%0\leq P(i,f) \leq 1, ~~ i\in \mathcal{N}, ~~ f\in \mathcal{F}   \label{cons1} \\
0 \leq \sum \nolimits_{i=1}^{N}P(i,f)\leq 1, ~\forall~f\in \mathcal{F}, \label{cons2} \\
\sum \nolimits _{f=1}^{F}P(i,f)S_{f} \leq M_i, ~\forall~i\in \mathcal{N}. \label{cons3}
\end{eqnarray}
\end{subequations}
Here (\ref{cons2}) ensures at most one copy of a content is cached in a cluster; (\ref{cons3}) guarantees the size of cached contents at each fog node cannot exceed its cache capacity; $S_{f}$ is the size of content $f$ and $M_i$ is the cache capacity of fog node $i$.
\end{definition}

\vspace{-0.6cm}

\subsection{Content Request and Provision}

\vspace{-0.2cm}

The popularity of content $f$, characterized by how frequent it is requested within a given time duration, is defined as $P_{r}(f)$, where $0< P_{r}(f) < 1$ and $\sum_{f=1}^{F}P_{r}(f) = 1$. Without loss of generality, the contents can be sorted in a popularity-descending order \cite{fogcaching_latency2, fog_caching4, segement}, i.e., $P_{r}(1)\geq P_{r}(2)\geq \cdots \geq P_{r}(F)$.

Upon receiving a request for content $f$, the associated BS will combine the segments related to content $f$ that are cached in the fog cluster, i.e., $\sum_{i=1}^{N}P(i,f)$, and then send them to the requesting user.

%We follow a commonly adopted setting and ignore the wired transmission delay among fog nodes.
%{\color{red} Compared with wireless transmission between fog node and users, the high-speed wired link among fog nodes can achieve much smaller transmission time, which will be ignored in the next analysis.}

\vspace{-0.2cm}

\begin{definition}\label{def2}
\vspace{-0.1cm}
Upon receiving a request for an arbitrary content $f$, we define ECHR as the average portion of content $f$ that can be found in the network edge, given by $H_{e}(\mathbf{P})=\sum \nolimits_{f=1}^{F}P_{r}(f)\left[\sum \nolimits_{i=1}^{N}P(i,f)\right]$.
%\begin{equation}\label{cache hit probability}
%H_{e}(\mathbf{P})=\sum \limits_{f=1}^{F}P_{r}(f)\left[\sum \nolimits_{i=1}^{N}P(i,f)\right].
%\end{equation}
\end{definition}
\vspace{-0.2cm}
For the remaining portion of content $f$, it has to be fetched from the CDC according to the corresponding backhaul-traffic-ratio (BTR) $H_{b}(\mathbf{P})=1-H_{e}(\mathbf{P})$.

%可以删除
%\begin{figure}[t]
%\begin{center}
%\vspace{-0.2cm}
%{\includegraphics[width=3in]{Fig2.eps}}
%\end{center}
%\vspace{-0.5cm}
%\caption{An illustration of the content request and provision at a fog node.}\label{system2}
%\vspace{-0.7cm}
%\end{figure}

\vspace{-0.3cm}

\section{Optimal ADT Analysis}

\vspace{-0.2cm}

\subsection{Problem Formulation}

\vspace{-0.2cm}

%As shown in Fig. \ref{system2},

It is assumed that the users' content requests arrived at BS $i\in \mathcal{N}$ follow a Poisson process with parameter $\lambda_{i}$. By employing a multi-class processor queue, the content provision can then be divided into two modes: \textit{fog provision mode} with ECHR $H_{e}(\mathbf{P})$ and \textit{cloud provision mode} with BTR $H_{b}(\mathbf{P})$  \cite{MPSQ}. Thus each mode can be modeled as an M/M/1 queuing process, with respective request arrival rates
\begin{equation} \label{arrival}
\lambda_{e,i}=\lambda_{i} H_{e}(\mathbf{P}) ~~  \text{and} ~~
\lambda_{b,i}=\lambda_{i} H_{b}(\mathbf{P}).
\end{equation}

%As shown in Fig. \ref{system2}, the users' content requests arrived at a fog node are assumed to follow a Poisson process with parameter $\lambda$. By employing a multi-class processor sharing queue \cite{MPSQ} at each fog node, the content provision can then be divided into two modes: \textit{fog provision} mode with relevant performance metric ECHR $H_{e}(\mathbf{P})$ and \textit{cloud provision} mode with relevant performance metric BTR $H_{b}(\mathbf{P})$ \cite{MPSQ}. We follow a commonly adopted setting and model each mode as a M/M/1 queue process, with request arrival rates
%\begin{equation} \label{arrival}
%\lambda_{e}=\lambda H_{e}(\mathbf{P})  \qquad \text{and} \qquad
%\lambda_{b}=\lambda H_{b}(\mathbf{P}).
%\end{equation}
% the associated BS $k$ will first collect the cached portion of the requested content in fog cluster and then send them to the requesting user with average content provision rate $\mu_{e,k}$. In cloud provision mode, the remaining portion of the requested content will need to be firstly fetched from the cloud to the associated BS $k$ and then delivered to the requesting user with average content provision rate $\mu_{b,k}$
%\begin{equation}
%\mu_{e,k}=R_{e,k}/S  ~~~ \text{and} ~~~
%\mu_{b,k}=1/(S/R_{e,k}+S/R_{b,k}),
%\end{equation}

In the \textit{fog provision mode}, the associated BS $i$ will first collect the cached portion of the requested content in the fog cluster and then send them to the requesting user with a pre-determined rate of $R_{e,i}$ \cite{bigdata_caching}. In the \textit{cloud provision mode}, the remaining portion of the requested content will be firstly fetched from the cloud to the associated BS $i$ with a rate of $R_{b,i}$ \cite{bigdata_caching}, and then delivered to the requesting user with rate $R_{e,i}$. Suppose that all contents have equal size $S$ \cite{cache_air}, \cite{fogcaching_latency2}, \cite{segement}, the corresponding content provision rates under these two modes are given by $\mu_{e,i}=R_{e,i}/S$ and $\mu_{b,i}=1/(S/R_{e,i}+S/R_{b,i})$ respectively, where we have $\lambda_{i}<\mu_{b,i}<\mu_{e,i}$ as the condition of stability \cite{Xiao_JSAC, admm}. Then the corresponding ADTs under the two modes are given by
\begin{equation}\label{ADT_eb}
T_{e,i}= 1/(\mu_{e,i}-\lambda_{e,i}) ~~~ \text{and} ~~~
T_{b,i}=1/(\mu_{b,i}-\lambda_{b,i}),
\end{equation}
respectively. Thus, the ADT $D_{i}$ at BS $i$ is expressed as
\begin{eqnarray} \label{Di}
%D_{i}=\frac{\lambda_{e,i}}{\lambda_{i}}T_{e,i}+\frac{\lambda_{e,i}}{\lambda_{i}}T_{b,i}=\frac{H_{e}(\mathbf{P})}{\mu_{e,i}-\lambda_{i} H_{e}(\mathbf{P})}+\frac{1-H_{e}(\mathbf{P})}{\mu_{b,i}-\lambda_{i} [1-H_{e}(\mathbf{P})]}
%D_{i}&=&\frac{\lambda_{e,i}T_{e,i}+\lambda_{b,i}T_{b,i}}{\lambda_{i}}, \nonumber \\
D_{i}=\frac{H_{e}(\mathbf{P})}{\mu_{e,i}-\lambda_{i} H_{e}(\mathbf{P})}+\frac{1-H_{e}(\mathbf{P})}{\mu_{b,i}-\lambda_{i} [1-H_{e}(\mathbf{P})]}. \label{ADT}
\end{eqnarray}

We define a performance metric $D=\sum_{i=1}^{N}\frac{\lambda_{i}}{\sum_{i=1}^{N}\lambda_{i}}D_{i}$ that reflects the overall ADT per request generated within the coverage of all BSs. Then subject to both the limited cache capacities of fog nodes and content provision rates of BSs, the problem of minimizing the overall ADT can be formulated as
\begin{eqnarray}\label{problem formulation}
\min \limits_{\mathbf{P}} D, ~~\mbox{s.t.}~(\ref{cons2});~(\ref{cons3});~\lambda_{i}<\mu_{b,i}<\mu_{e,i}~\forall~i\in \mathcal{N}.
\end{eqnarray}

\vspace{-0.4cm}
\subsection{Algorithm Design}
\vspace{-0.1cm}
%the cache content placement matrix $\mathbf{P}$
%$\boldsymbol{\beta} = [P(1,1),..., P(1,F),..., P(N,1),..., P(N,F)]^{T}$
\begin{theorem}
Based on the cache content placement matrix $\mathbf{P}$ defined in \textbf{Definition 1}, we define the cache content placement vector $\mathbf{p}=\langle P(i,f) \rangle_{i\in \mathcal{N},f\in \mathcal{F}}$. Then the problem in (\ref{problem formulation}) is a convex optimization problem with respect to $\mathbf{p}$.
\end{theorem}
\vspace{-0.3cm}
\begin{proof} See Appendix \ref{appendixA}.\end{proof}

To solve the problem in (\ref{problem formulation}), we propose an algorithm based on ADMM, which is applicable to convex optimization problems in distributed deployment scenarios with fast convergence rate \cite{admm}. However, basic ADMM cannot be directly applied to our problem in (\ref{problem formulation}) with inequality constraints \cite{Xiao_JSAC, admm}. To solve this issue, we reformulate (\ref{problem formulation}) as the following form
\begin{eqnarray}
&\min \limits_{\mathbf{p}} D(\mathbf{p}) \nonumber\\
&\mbox{s.t.}~ 0 \leq \mathbf{p}(j) \leq 1, 1\leq j \leq{NF}, ~\mathbf{A} \mathbf{p} \leq \mathbf{A_{u}}, \mathbf{B} \mathbf{p} \leq \mathbf{B_{u}}, \label{admm_problem1}
\end{eqnarray}
where $\mathbf{A_{u}} = [1, \cdots, 1]^{T} \in \mathbb{R}^{F\times 1}$ denotes the upper bound of (\ref{cons2}), $\mathbf{B_{u}} = [M_1, \cdots, M_N]^{T} \in \mathbb{R}^{N\times 1}$ denotes the upper bound of (\ref{cons3}), and matrices $\mathbf{A}\in \mathbb{R}^{F \times NF}$ and $\mathbf{B}\in \mathbb{R}^{N \times NF}$ are block-diagonal matrices that can get the summation of sub-vectors of $\mathbf{p}$ corresponding to (\ref{cons2}) and (\ref{cons3}).

%denote the operations of selecting the sub-vector of $\boldsymbol{\beta}$ corresponding to $\mathbf{A1}$ and $\mathbf{B1}$ respectively. Let $\mathbf{A}_{v_{1},w}$, where $1\leq v_{1}\leq F$ and $1\leq w \leq NF$, be the element in the $v_{1}$-row and $w$-column in $\mathbf{A}$, we have
%\begin{eqnarray}
%\mathbf{A}_{v_{1},w}=\begin{cases}
%1, & \text{if}~~ w = v_{1} + F\cdot k, ~~ k=0,1,2,...  ,\\
%0, & \text{otherwise}.
%\end{cases}
%\end{eqnarray}
%Similarly for $\mathbf{B}_{v_{2},w}$, where $1\leq v_{2}\leq N$, we have
%\begin{eqnarray}
%\mathbf{B}_{v_{2},w}=\begin{cases}
%1, & \text{if}~~ v_{2} \leq w < (v_{2}+F) , \\
%0, & \text{otherwise}.
%\end{cases}
%\end{eqnarray}
%\begin{eqnarray}
%&g(\mathbf{z}) =
%\begin{cases}
%0, & \mathbf{z} \in \mathcal{C} ,\\
%+\infty, & \mathbf{z} \notin \mathcal{C} ,\\
%\end{cases}
%\end{eqnarray}

We introduce an indicator function $g(\mathbf{z})$ defined on $\mathcal{C}=\left\{\mathbf{z}:\forall j, 0 \leq \mathbf{z}(j) \leq 1, \mathbf{A} \mathbf{z} \leq \mathbf{A_{u}}, \mathbf{B} \mathbf{z} \leq \mathbf{B_{u}}\right\}$ for the inequality constraints in (\ref{admm_problem1}), where $\mathbf{z}\in \mathbb{R}^{NF \times 1}$ is the auxiliary vector. Then problem (\ref{admm_problem1}) can be further reformulated as
\begin{eqnarray}\label{admm_problem2}
\min \limits_{\mathbf{p}} D(\mathbf{p})+g(\mathbf{z}) ~~~~~~\mbox{s.t.}~~~ \mathbf{p}-\mathbf{z} = 0.
\end{eqnarray}

We can solve (\ref{admm_problem2}) via the following updating scheme \cite{admm}:
\begin{subequations}
\begin{eqnarray}
\mathbf{p}^{k+1} &=& \arg\min \limits_{\mathbf{p}} \left(D(\mathbf{p})+\frac{\rho}{2}\parallel \mathbf{p}-\mathbf{z}+\boldsymbol{\theta} \parallel _{2}^{2} \right), \label{psi} \\
\mathbf{z}^{k+1} &=& \Pi _\mathcal{C}(\mathbf{p}^{k+1}+ \boldsymbol{\theta}^{k}), \label{z} \\
\boldsymbol{\theta}^{k+1} &=& \boldsymbol{\theta}^{k}+\mathbf{p}^{k+1}-\mathbf{z}^{k+1}, \label{mu}
\end{eqnarray}
\end{subequations}
where $\rho$ is the augmented Lagrangian factor and $\boldsymbol{\theta} \in  \mathbb{R}^{NF \times 1}$ is the dual variable \cite{admm}.

We present the details of our proposed cache content placement algorithm in $\mathbf{Algorithm ~\ref{ADMMalg}}$. Following the same line as \cite{admm}, we can prove that $\mathbf{Algorithm ~\ref{ADMMalg}}$ can converge to the global optimal solution to (\ref{admm_problem1}) at a rate of $\mathcal{O}(1/t)$. The details are omitted here due to space limit.

\begin{algorithm}[t]
\caption{:~Optimal Cache Content Placement Algorithm.}
\label{ADMMalg}
%\begin{spacing}{1.0}
\begin{algorithmic}
\STATE a)~Initialization: The system chooses an initial cache content placement vector $\mathbf{p}^{0}$, slack variable $\mathbf{z}^{0}$ and dual variable $\boldsymbol{\theta}^{0}$.
\STATE b)~At $(k+1)$-th iteration:
\begin{enumerate}
\item Cache content placement vector update: calculate $\mathbf{p}^{k+1}$ based on (\ref{psi});\\
\item Slack variable update: calculate $\mathbf{z}^{k+1}$ based on (\ref{z});\\
\item Dual variable update: calculate $\boldsymbol{\theta}^{k+1}$ based on (\ref{mu});\\
\end{enumerate}
\STATE c)~If the constraint is violated, increase $k$ and go to step b. Otherwise, all fog nodes cooperatively execute the cache content placement based on the obtained vector $\mathbf{p}^{k+1}$.
\end{algorithmic}
\end{algorithm}
\vspace{-0.2cm}

\begin{figure*}[t]
\centering
\vspace{-0.2cm}
\subfigure[Comparison of convergence rate.]{
\includegraphics[width=1.7in]{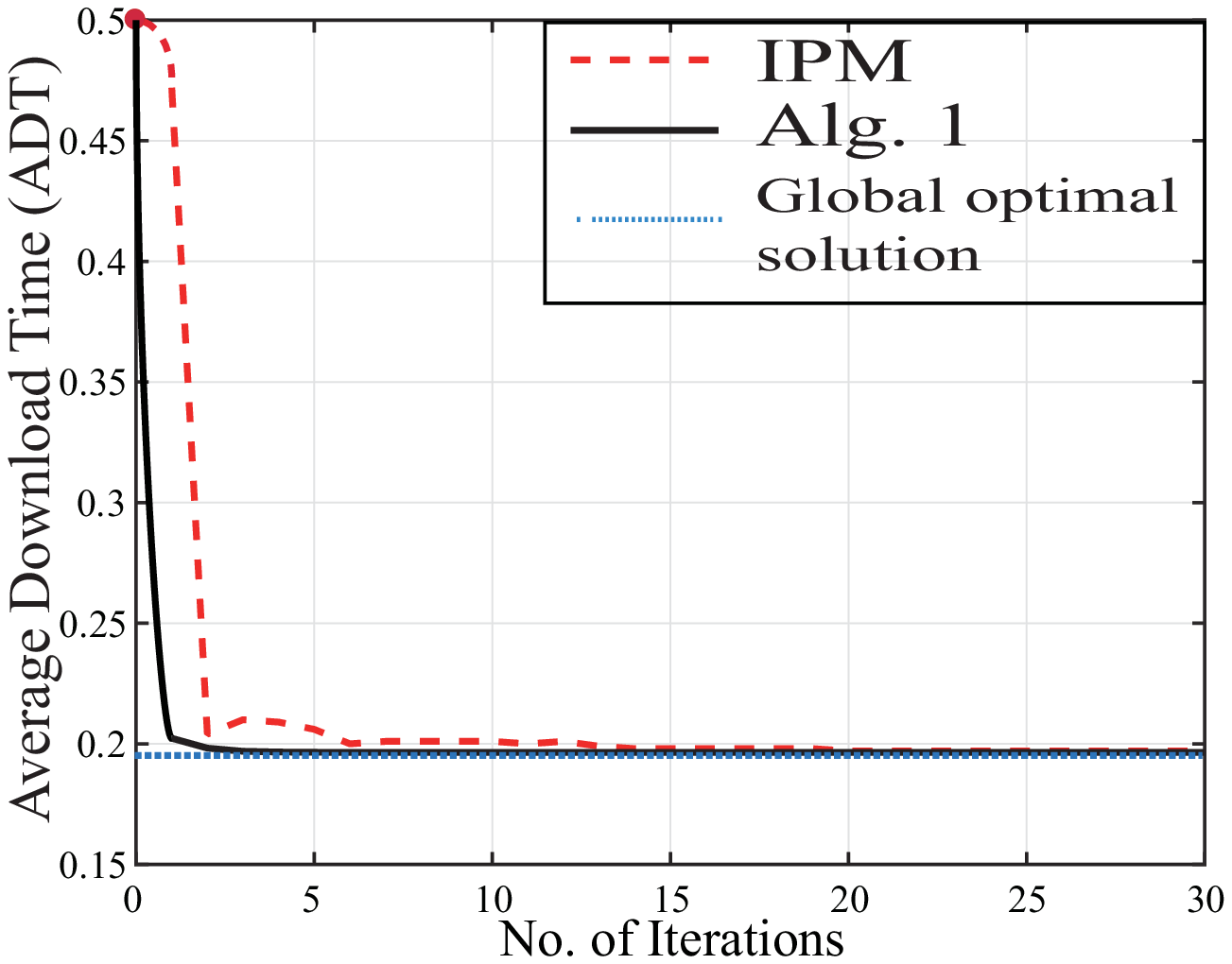}\label{iteration}}
\subfigure[ECHR when $F=18$.]{
\includegraphics[width=1.7in]{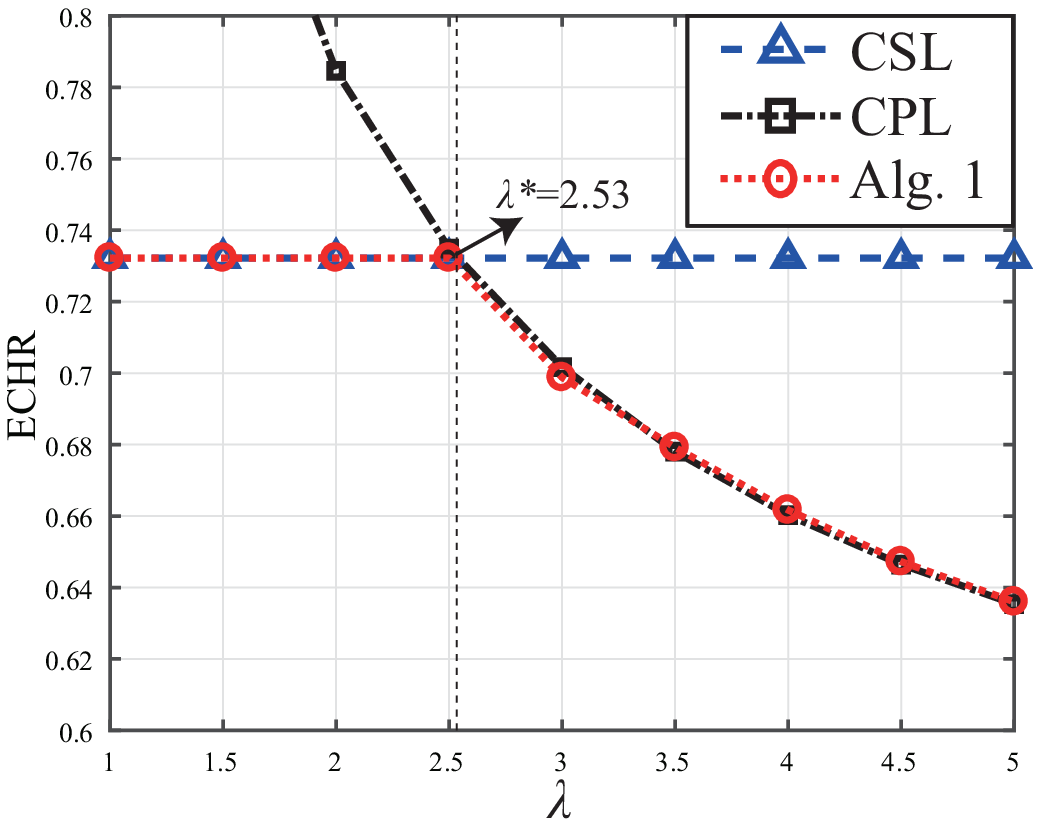}\label{sim_ECHR_1}}
\subfigure[ECHR when $F=20$.]{
\includegraphics[width=1.7in]{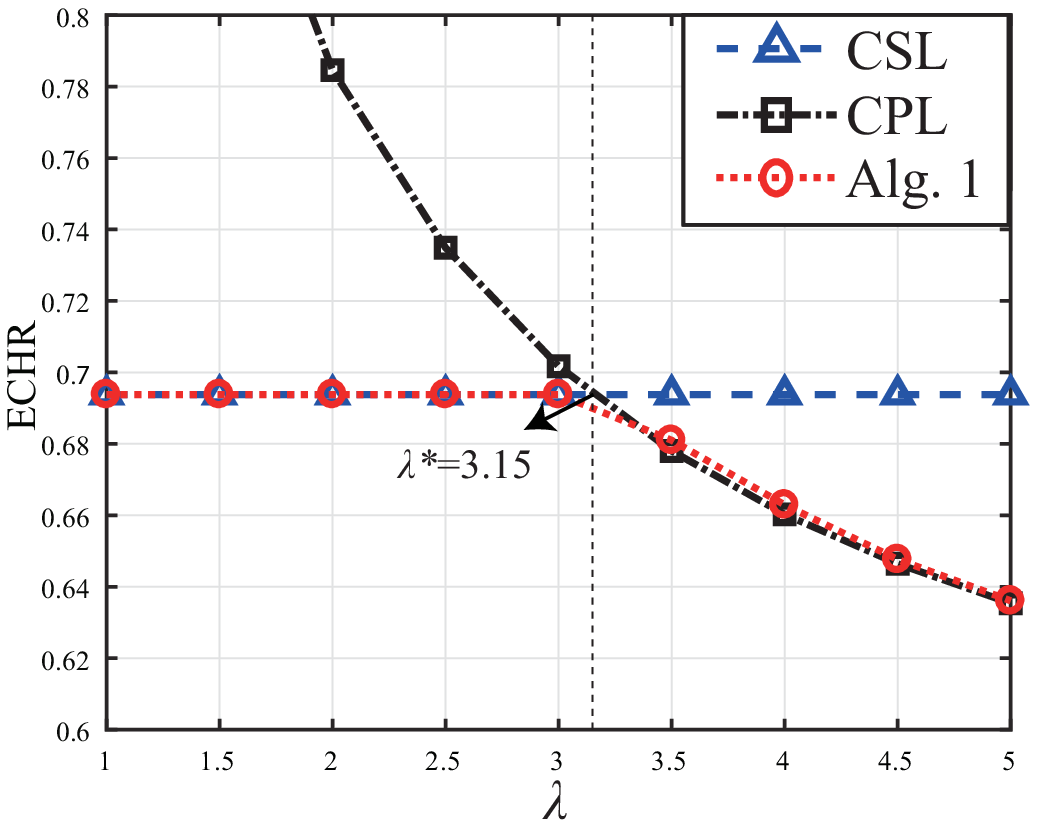}\label{sim_ECHR_2}}
\subfigure[ECHR when $F=22$.]{
\includegraphics[width=1.7in]{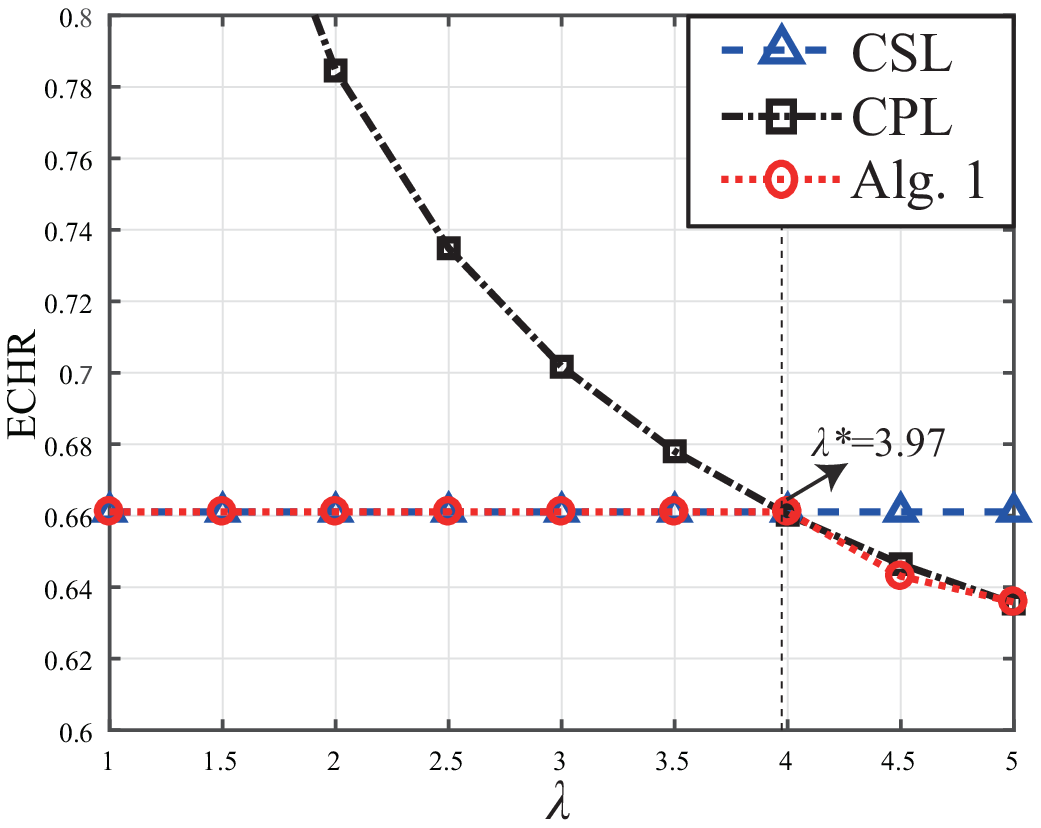}\label{sim_ECHR_3}}
\subfigure[ADT and ECHR with respect to $\mu_b$.]{
\includegraphics[width=2in]{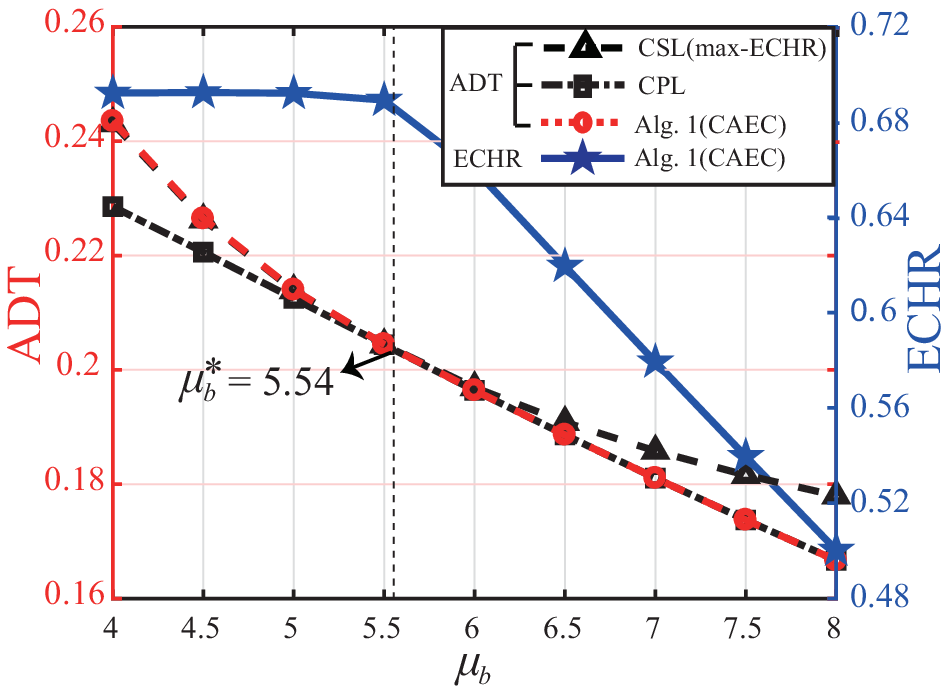}\label{ADT_mub}}
\subfigure[ADT and ECHR with respect to $\lambda$.]{
\includegraphics[width=2in]{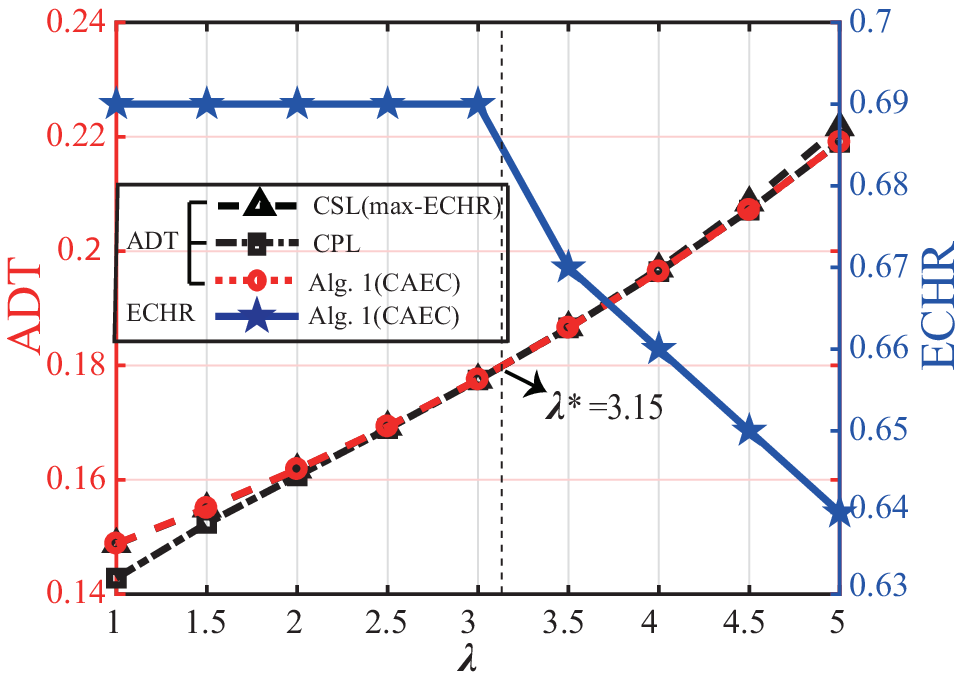}\label{ADT_lambda}}
\subfigure[ADT and ECHR with respect to $\mu_e$.]{
\includegraphics[width=2in]{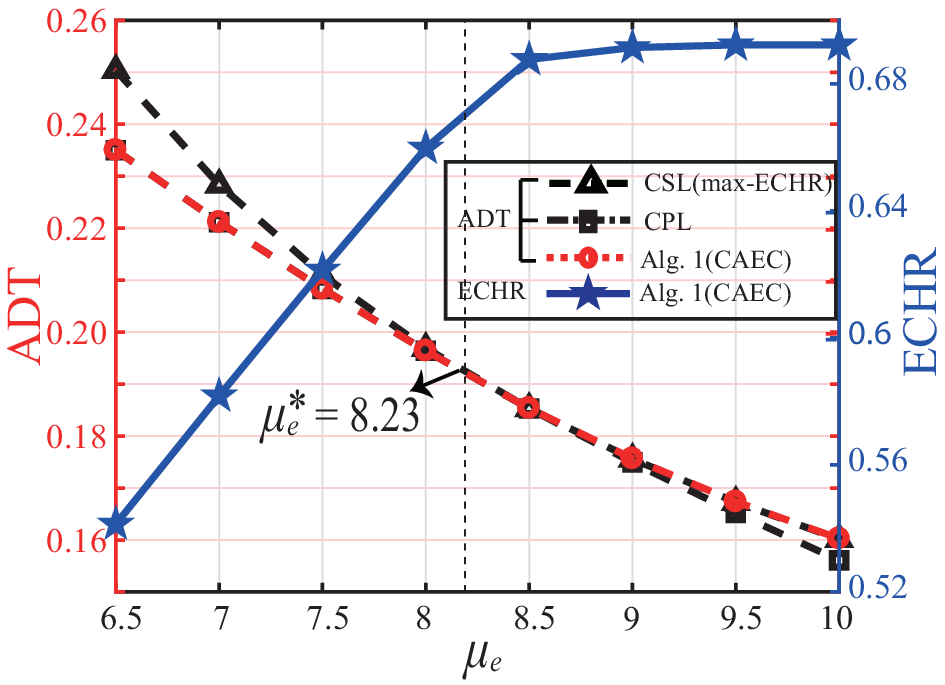}\label{ADT_mue}}
\caption{An illustration of the optimal ECHR and ADT under different cases.}\label{sim}
\vspace{-0.6cm}
\end{figure*}

\vspace{-0.1cm}

\subsection{A Heuristic Caching Method}
\vspace{-0.2cm}

To provide insights into the relationship between the constraints of cache capacities of fog nodes and the content provision rates caused by limited BS connectivity capacities, we further propose a simple heuristic caching method based on the following two special cases.

\subsubsection{Cache-Storage-Limited (CSL) Case}
In this case, the content provision rate is sufficiently high such that a cached content can be provisioned instantaneously upon a request, however only a limited number of contents can be stored due to the limited cache capacity, which causes the bottleneck of the ADT. As a result, minimizing $D$ is equivalent to maximizing $H_{e}(\mathbf{P})$ in \textbf{Definition}~\ref{def2} subject to (\ref{cons2}), (\ref{cons3}). Then the optimal $\mathbf{P}$ for achieving the maximum ECHR, i.e., $H_{e}^{\textmd{csl}}$, can be obtained by using the Interior Point Method (IPM) \cite{convex}, as shown in Section IV.

%, e.g., dedicated short range communications based fog nodes

%We consider the conventional CSL case where the content provision rate at each fog node is bottlenecked by the cache \cite{cache_air, fogcaching_maxhit, D2D_ECHR, fogcaching_latency1, fogcaching_latency2}. Then the minimization of $D$ is equivalent to the maximization of the ECHR $H_{e}$ defined in (\ref{cache hit probability}) subject to constraints (\ref{cons1})-(\ref{cons3}) \cite{delay_echr}. By using the IPM \cite{convex}, the optimal cache content placement matrix $\mathbf{P}$ for achieving the maximum ECHR, i.e., $H_{e}^{\textmd{csl}}$, can be obtained. The details are omitted due to space limit.

\subsubsection{Content-Provision-Limited (CPL) Case}
In this case, fog cache capacity is sufficiently large such that almost all contents can be stored. However, the cached contents cannot be promptly provisioned due to the limited content provision rate, which causes the bottleneck of the ADT. According to Appendix A, since $D$ is convex with respect to $H_{e}(\mathbf{P})$, by setting $\frac{\partial D}{\partial H_{e}(\mathbf{P})}=0$, we can obtain the optimal ECHR $H_{e}^{\textmd{cpl}}$ under the CPL case.

Since both constraints on fog cache capacity and content provision rate must be met, we propose a heuristic caching method by switching between CSL and CPL, with the corresponding ECHR $H_{e}^{\ast}=\min\{H_{e}^{\textmd{csl}},H_{e}^{\textmd{cpl}}\}$.

Specifically, when each BS experiences homogeneous traffic pattern that $\lambda_i=\lambda,\mu_{e,i}=\mu_e,\mu_{b,i}=\mu_b~\forall~i\in \mathcal{N}$, we obtain
\begin{eqnarray}\label{solution_he}
H_{e}^{\ast}=\begin{cases}
H_{e}^{\textmd{csl}}, & \text{if}~~ \lambda < \lambda^{\ast}  \\
H_{e}^{\textmd{cpl}}, & \text{otherwise}
\end{cases},
\end{eqnarray}
where $H_{e}^{\textmd{cpl}}=\frac{(\mu_{e}-\sqrt{\mu_{e} \mu_{b}})\sqrt{\mu_{b}}+\lambda \sqrt{\mu_{e}}}{\lambda \sqrt{\mu_{b}}+\lambda \sqrt{\mu_{e}}}$ and $\lambda^{\ast}=\frac{\sqrt{\mu_{b}\mu_{e}}(\sqrt{\mu_{e}}-\sqrt{\mu_{b}})}{H_{e}^{\textmd{csl}}(\sqrt{\mu_{e}}+\sqrt{\mu_{b}})-\sqrt{\mu_{e}}}$ is obtained by letting $H_{e}^{\textmd{csl}}=H_{e}^{\textmd{cpl}}$.

%缩减/删
\begin{remark}
From (\ref{solution_he}), it can be observed that the proposed heuristic caching method corresponds to CSL (or CPL) in low (or heavy) traffic where $\lambda < \lambda^{\ast}$ (or $\lambda >\lambda^{\ast}$). Since a mismatch between the fog cache capacity and content provision rate may result in an under-utilization of available storage and communication resources, these two factors need to be balanced to provide satisfactory services and at the same time utilize the resources efficiently.
\end{remark}

\vspace{-0.45cm}

\section{Simulation Results}

\vspace{-0.1cm}
% contents respectively
In this section, we present simulation results to evaluate the performance of the proposed CAEC. For ease of illustration, we let $F=20$ and assume the content popularity follows a Zipf distribution with parameter $\alpha = 0.6$ \cite{fogcaching_latency2}, \cite{fog_caching4}, \cite{segement}. We consider a cluster of $N=3$ fog nodes with cache capacities $[2S;3S;5S]$, and let $\lambda_i = \lambda= 4$, $\mu_{e,i}=\mu_{e}=8$ and $\mu_{b,i} = \mu_{b}= 6~\forall~i\in \{1,2,3\}$, unless otherwise specified.

From Fig. \ref{iteration}, it is observed that while both IPM and $\mathbf{Algorithm ~\ref{ADMMalg}}$ can converge to the same global optimal ADT, $\mathbf{Algorithm ~\ref{ADMMalg}}$ converges much faster (less than 5 iterations).

%删减
The optimal ECHR is demonstrated in Fig. \ref{sim_ECHR_1}-Fig. \ref{sim_ECHR_3} with respect to $\lambda$. It is observed that while $H_{e}^{\textmd{csl}}$ depends on the available cache capacity only and is irrelevant to $\lambda$, $H_{e}^{\textmd{cpl}}$ decreases with $\lambda$. This is because with a higher $\lambda$, more requests should be allocated to the cloud for keeping a lower ECHR in the CPL case, such that the cached content can be promptly provisioned. Furthermore, we observe that the proposed heuristic caching method in (\ref{solution_he}), with a switch between CSL and CPL at $\lambda^{\ast}$, achieves an ECHR that is very close to the global optimal ECHR obtained by $\mathbf{Algorithm ~\ref{ADMMalg}}$.

%Similar results can be observed in Fig. \ref{sim_ECHR_2} and Fig. \ref{sim_ECHR_3}, except that an overall lower ECHR is achieved with increasing $F$.

%删减
In Fig. \ref{ADT_mub}-Fig. \ref{ADT_mue}, we present the optimal ADT and ECHR with respect to $\mu_b$, $\lambda$ and $\mu_e$, respectively. It is observed from Fig. \ref{ADT_mub} that the overall ADT performance improves with $\mu_b$. In the low-$\mu_b$ regime, it tends to select a high ECHR to avoid waiting too long in the cloud provision queue, i.e., the CSL case. Furthermore, in the high-$\mu_b$ regime, it tends to select a lower ECHR to avoid long queueing delay in the fog provision queue, i.e., the CPL case. Similar result can be observed in Fig. \ref{ADT_lambda}, where the overall ADT performance degrades with increasing $\lambda$. Compared to Fig. \ref{ADT_mub}, opposite result can be observed in Fig. \ref{ADT_mue}, where the optimal ECHR increases with $\mu_e$, and the optimal ADT is limited by CPL and CSL in low-$\mu_e$ and high-$\mu_e$ regimes, respectively.

%Again, the proposed heuristic caching algorithm achieves a very close ADT to $\mathbf{Algorithm ~\ref{ADMMalg}}$.
%Again, it is observed that the proposed heuristic caching algorithm, with a switch between the CSL and CPL at $\mu^{\ast}_b=5.54$, achieves a very close ADT to $\mathbf{Algorithm ~\ref{ADMMalg}}$.
%In the low-$\lambda$ regime, the ADT is limited by the limited cache (i.e., the CSL case), whereas in the high-$\lambda$ regime, the ADT is limited by the limited service processing rate (i.e., the PSL case).

\vspace{-0.2cm}

\section{Conclusion}
\vspace{-0.2cm}
In this paper, a CAEC framework is proposed for efficient content cache and provision in fog computing networks. By jointly considering the limited cache capacities of fog nodes and BS connectivity capacities, the minimization of ADT per arbitrary requested content is formulated as a multi-class processor queue problem. An algorithm based on ADMM is proposed to efficiently search the optimal cache content placement at each fog node. Furthermore, a simple heuristic caching algorithm is proposed to achieve a near-optimal performance of the proposed algorithm. Simulation results demonstrate that maximizing ECHR is not the optimal strategy under certain scenarios and the proposed CAEC scheme can always provide a more balanced traffic allocation strategy that has the minimum ADT.

%%future workx
%unknown time-static popularity distribution
%time-varying popularity distribution
%asymmetric traffic load-> popularity distribution
%heterogeneous user preference

%{D_i}  = \frac{{{H_e}\left( {\mathbf{P}} \right)}}{{{\mu _{e,i}} - {\lambda _i}{H_e}\left( {\mathbf{P}} \right)}} + \frac{{1 - {H_e}\left( {\mathbf{P}} \right)}}{{{\mu _{b,i}} - {\lambda _i}\left( {1 - {H_e}\left( {\mathbf{P}} \right)} \right)}}\\
%\frac{{\partial {D_i}}}{{\partial {H_e}\left( {\mathbf{P}} \right)}} = \frac{{{\mu _{e,i}}}}{{{{\left( {{\mu _{e,i}} - {\lambda _i}{H_e}\left( {\rm{P}} \right)} \right)}^2}}} + \frac{{ - {\mu _{b,i}}}}{{{{\left[ {{\mu _{b,i}} - {\lambda _i}\left( {1 - {H_e}\left( {\rm{P}} \right)} \right)} \right]}^2}}}\\

\vspace{-0.2cm}

\appendices
\section{Proof of Theorem 1}\label{appendixA}
%\vspace{-0.1cm}
From (\ref{Di}), we have $\frac{{\partial D_i^2}}{{{\partial ^2}{H_e}\left( {\mathbf{P}} \right)}} = \frac{{2{\mu _{e,i}}{\lambda _i}}}{{{{\left( {{\mu _{e,i}} - {\lambda _i}{H_e}\left( {\mathbf{P}} \right)} \right)}^3}}} + \frac{{2{\mu _{b,i}}{\lambda _i}}}{{{{\left[ {{\mu _{b,i}} - {\lambda _i}\left( {1 - {H_e}\left( {\mathbf{P}} \right)} \right)} \right]}^3}}}$. Since $0 < {\lambda _i} < {\mu _{b,i}} < {\mu _{e,i}}$ and $0\leq H_{e}(\mathbf{P})\leq1$, we have $\frac{{\partial D_i^2}}{{{\partial ^2}{H_e}\left( {\mathbf{P}} \right)}} > 0, ~\forall i \in \mathcal{N}$. Furthermore, let $\gamma = \left[ {{P_r}(1),{P_r}(2),...,{P_r}(F)} \right]$, we have ${H_e}\left( \mathbf{P} \right) = \gamma {\mathbf{A}} \mathbf{p}$, $\frac{{\partial^{2} H_e \left( {\mathbf{P}} \right)}}{{\partial^{2} {\mathbf{p}}}}{\rm{ = }} \mathbf{0}$. Thus, the Hessian matrix of $D_{i}$ is $\mathbf{H}=\frac{{\partial D_i^2}}{{{\partial ^2}\mathbf{p} }}
=\frac{{\partial D_i^2}}{{{\partial ^2}{H_e}\left( {\mathbf{P}} \right)}} (\gamma \mathbf{A})^{T} (\gamma \mathbf{A})$, and it can be derived that  $\mbox{det}(\mathbf{H})=0$ and $\mathbf{H}$ is a non-negative semi-definite matrix. Since $D=\sum_{i = 1}^N \frac{\lambda_{i}}{\sum\nolimits_{i = 1}^N \lambda_{i}} D_i$, $\mbox{det}(\frac{{\partial {D^2}}}{{{\partial ^2}\mathbf{p} }})=0$ and the Hessian matrix of $D$ is also a non-negative semi-definite matrix. Thus \textbf{Theorem 1} is proved.

%$\frac{{\partial {D^2}}}{{{\partial ^2}\mathbf{p} }}{\rm{ = }}\sum_{k = 1}^K \left[\frac{\lambda_{k}}{\sum\nolimits_{k = 1}^K \lambda_{k}}  {\frac{{\partial D_k^2}}{{{\partial ^2}\mathbf{p} }}}  \right] = 0$
%$\frac{{\partial H_e \left( {\mathbf{P}} \right)}}{{\partial {\mathbf{p}}}}{\rm{ = }} (\gamma \mathbf{A})^{T}$ and

\end{document}